# New perspectives on the fluorite – pyrochlore phase transition in $La_2Zr_2O_7$ and the importance of local oxygen-related disordered states

Barnita Paul,[‡] Kushal Singh,[†] Tomasz Jaroń,[|] Anushree Roy,[‡] Anirban Chowdhury,[*,†]

[‡] Department of Physics, Indian Institute of Technology, Kharagpur, 721302. India

[†] Department of Materials Science and Engineering, Indian Institute of Technology Patna 80013, Bihar, India

[|] Centre of New Technologies, University of Warsaw, ul. Zwirki i Wigury 93, 02-089 Warsaw, Poland

**Abstract**

The fluorite to pyrochlore phase transition in $La_2Zr_2O_7$ has been studied in the literature for decades in the context of thermal barrier coatings and reinforcement materials. However, the nature of the phase transition in this system is still not well understood. In this article we have investigated the phase transition in $La_2Zr_2O_7$, calcined at different temperatures, using powder x-ray diffraction and Raman measurements. Rietveld analyses of the x-ray data suggest a fluorite to pyrochlore phase transition in the system occurring between 1000-1450$^o$C. Nonetheless, Raman spectra, recorded with different excitation wavelengths ($\lambda_{ex}$) reveal that the dynamics of phase transition is different in the near-surface region and in the bulk inside. While the near-surface region carried the signatures of a pure pyrochlore phase, the temperature dependent Raman measurements with $\lambda_{ex}$=785 nm suggest that locally formed oxygen related disordered states are present in the bulk of the system.

**Introduction**

Lanthanum zirconate, $La_2Zr_2O_7$ (LZ) is a technologically important material due to its low thermal conductivity and superior high-temperature stability.[1] This compound has found its applications in zirconia-based thermal barrier coatings and as a reinforcement material.[2,3] Other areas of application include the field of superconductors where $La_2Zr_2O_7$ is used as a buffer layer.[4,5] In $A_2B_2O_7$ type structure, the pyrochlore phase is stable when the ratio of ionic radii of A to B cations is above 1.26. Below this value of the ionic radii the system is expected to remain in the fluorite phase.[6] For $La_2Zr_2O_7$ the value of this ratio is 1.40, which is not far from the above phase determining value of the ratio 1.26. Thus, it is expected that the structural phase transitions of $La_2Zr_2O_7$ will strongly depend on the inherent disorder in the crystal structure following different growth conditions. It has always been a challenge for the scientific community, at large, to study the origin of structural phase transition in the system under pressure or temperature.[7,8]

The fluorite structure in $La_2Zr_2O_7$ originates from $CaF_2$ displaying a cubic structure with Fm3m space group (No. 225). The disordered fluorite structure of same space group can be explained from the cubic $ZrO_2$ lattice with $La^{3+}$ ions and $Zr^{4+}$ ions randomly distributed on $Zr^{4+}$ sites and $7/8^{th}$ occupancy of O atoms in the anionic sublattice.[9] The pyrochlore structure ($A_2B_2O_6$ and $A_2B_2O_7$) is a super-structure derivative of the simple fluorite structure where the A and B cations are ordered along the <110> direction. The additional anion vacancy resides in the tetrahedral interstice between adjacent B-site cations.[10]

A high-temperature order-disorder phase transformation has been commonly observed for $A_2B_2O_7$ type rare-earth oxide systems at elevated temperatures (1530–2400°C).[8,11,12] $Ln_2Ti_2O_7$ (Ln= Tm-Lu), and $Gd_2M_2O_7$ (M = Zr, Hf) compounds prepared by co-precipitation showed low-temperature (800 –1000 °C) fluorite to pyrochlore transformations.[13-16] In the case of $La_2Zr_2O_7$ contradicting reports on order-disorder phase transition can be found in the literature. In some of the reports the fluorite-type phase is claimed to be stable at high temperature but upon cooling it transforms into pyrochlore phase with an ordering of cations and anionic vacancies[13]. Other sources mention that there is a fluorite to pyrochlore phase transition in this system above 1000 °C.[17] However, Michel *et al.*[8, 12] claimed that there is no high-temperature fluorite-pyrochlore transition until the melting point. In contrary, the phase transition could be observed in $Ln_2Zr_2O_7$ structures (Ln= Nd, Sm, Gd) without melting of the system.[8] Radha et al[18] also concluded in the same line in their recent work. The thermodynamics of such low temperature (800–1100 °C) phase transformation is not well understood.[18] Amid of these conflicting reports, in a recent study[19] by one of the authors (AC) of this article has shown the role of non-stoichiometry of the compound in determining the fluorite or pyrochlore phase of this system.

To look into the microscopic origin of the phase transition further in detail, in this article we investigate the phase transition in $La_2Zr_2O_7$ calcined at different temperatures. While we exploit the sensitivity of the X-ray diffraction to find the long-range ordering in the crystal structure, the local atomic arrangement could be probed by Raman measurements. Furthermore, in Raman spectroscopy we take the advantage of the inverse relation of excitation frequency of laser radiation with its penetration depth in a material to probe and find local oxygen related disordered states in the bulk inside, which are absent in the near-surface region of the high temperature pyrochlore phase. We have shown that the disordered states can be annealed out upon cooling. However, they reappear under ambient conditions.

**Experiments**

To prepare Lanthanum zirconate powder, we have followed co-precipitation method using lanthanum nitrate and zirconium oxychloride as precursors. The details of the synthesis route have been discussed elsewhere.[19] The phase purity and stoichiometry issues mentioned in the previous work were addressed and resolved by carefully controlled washing by combining the centrifugation and ultrasonication steps together. The as-prepared powder was calcined at various temperatures ($T_c$): 750, 850, 950, 1000, 1100 and 1200 and 1450 °C. We will refer to these samples as LZ-750, LZ-850, LZ-950, LZ-1000, LZ-1100, LZ-1200 and LZ-1450, respectively.

The phase analysis of the lanthanum zirconate powders has been performed using powder diffraction patterns measured on Bruker AXS C-8 advanced X-ray diffractometer in Bragg-Brentano geometry. Cu-K$_{\alpha 1}$ and Cu-K$_{\alpha 2}$ radiation of intensity ratio *ca.* 2:1 was used. The diffraction angle (2θ) range was 10º - 90º, with a step size of 0.01–0.02º and scan time for each step of 10 s. This particular slow scan setting was used for data collection for Rietveld analysis, while the faster scans were performed to check the phase evolutions with respect to different $T_c$. The operating X-ray source voltage and current conditions were 50 kV and 100 mA, respectively. Rietveld refinement has been performed in Jana2006.[20] The background has been fitted by 20 Legendre polynomials and the peak shape has been modeled by pseudo-Voigt functions. Berar-Baldinozzi correction for the peak asymmetry and Pitschke, Mattern and Hermann correction for surface roughness and porosity have been applied.

Raman measurements with the excitation wavelength ($\lambda_{ex}$) =488 nm of Ar$^+$-Kr$^+$ (model 2018-RM, Newport, USA) were carried out using a micro-Raman spectrometer (model T64000, JY, France) equipped with Peltier cooled CCD detector (model-Synapse, JY, France). The same with the $\lambda_{ex}$=785 nm were carried out using a frequency-stabilized diode laser (model PI-ECL-785-300-FC, Process Instruments, Inc. USA). Spectra were recorded using a micro-Raman spectrometer (model- iHR 550, JY, France) equipped with Peltier cooled CCD (model Synapse, Horiba JY, USA). In both cases spectra are recorded in back-scattering geometry using a 50L× microscope objective lens. For temperature dependent

Raman measurements of LZ-1450 sample with 785 nm laser excitation, we have used a sample stage (Model THMS600, Linkam Scientific Instruments, UK) and a temperature controller. We have used 3 mW and 6 mW laser power on the sample surface for $\lambda_{ex}$= 488 nm and $\lambda_{ex}$= 785 nm, respectively.

**Results**

Due to clear structural similarities, it is difficult to distinguish the powder XRD patterns of the fluorite and pyrochlore phases of lanthanum zirconate. In addition to fluorite related peaks only a few minor additional (hkl) reflections are observed for pyrochlore phase (JCPDS file no. 01-071-2363), *e.g.*(111), (331), (511) and (800) at 14.21° (3.5%), 36.28° (4.7%), and 43.57° (1.8%) and 69.69 ° (4.4%), respectively. The powder XRD patterns of the calcined samples are shown in Figure 1. The XRD pattern of LZ-750 (not shown here) reveals a few very broad signals, indicating short range ordering is present in this system. The FWHM of the peaks gradually decreases with increase in $T_c$, which reveals the formation of crystalline phases along with considerable crystallite growth and agglomeration in the systems. The XRD scans of the samples calcined at 1000 ºC and above (LZ-1000 – LZ-1450) clearly shows the reflections confirming the presence of a pyrochlore-type phase by the appearance of above mentioned specific pyrochlore phase-related peaks.

As the strong diffraction peaks of lanthanum zirconate in fluorite and pyrochlore structure overlap, a careful examination is needed to interpret the measured XRD patterns. Therefore, the Rietveld refinement was carried out using good-quality diffraction patterns for the samples LZ-1000 and LZ-1450, Figure 2, Table 1. While in the sample LZ-1000 both pyrochlore- and fluorite-type phases are present in comparable amounts (*ca.* 54 : 46, respectively), the XRD pattern of the sample LZ-1450 is properly refined using the pyrochlore-type phase exclusively. Thus, according to the XRD results, the sample LZ-1450 is composed only of the pyrochlore-type phase; additionally a few very weak XRD signals from unidentified impurities (*ca.* 1% intensity of the strongest $La_2Zr_2O_7$ peak) can be detected in some of the prepared batches. The corresponding unit cell constants and interatomic distances in pyrochlore-type phase present in both LZ-1000 and LZ-1450 samples remain unchanged within the estimated standard deviations.

The strongly overlapping diffraction peaks of the fluorite and pyrochlore-type structures can seriously affect the structural parameters in Table 1 and especially the fractions of the two phases in the system. Thus, the structural phase transition in LZ systems, calcined at different temperatures, is further studied by Raman measurements. As mentioned earlier, a perfect cubic fluorite structure belongs to $Fm\bar{3}m$ space group. This space group has only one Raman active $F_{2g}$ mode, which involves oxygen ion vibration in a tetrahedral environment, formed by four *A*-cations. In an ideal cubic pyrochlore structure ($A_2B_2O1_{(6)}O2$) with $Fd\bar{3}m$ space group, one observes six Raman active modes. Five of them involve cation-anion vibration and the highest wavenumber mode is related only to oxygen sublattice. The vibrational normal modes at the center of the Brillouin zone are distributed on the irreducible representations as[21]

$$\Gamma_{pyro} = A_g^R + E_g^R + 7F_{1u}^{IR} + 4F_{2g}^R.$$

R and IR correspond to Raman and IR active modes. Figure 3 shows room temperature Raman spectra of LZ systems calcined at different temperatures. The spectra are recorded using 488 nm laser light as an excitation source. All spectra are baseline corrected after extracting in the range of 100–700 cm$^{-1}$. For LZ-750, the broad spectral feature centred at 350 cm$^{-1}$ corresponds to single Raman active mode of the disordered fluorite phase.[22] The other broad feature at 450 cm$^{-1}$ may arise due to oxygen non-stoichiometry in the parent compound .[23] The broad spectral profile indicates that in LZ-750 anions are coordinated in a random manner in the fluorite phase, introducing a short range structural disorder in the atomic arrangement in the lattice. With an increase in $T_c$ the Raman spectral profile becomes sharper.

The Raman lines of LZ-1450 (top panel of Figure 3) is found to be sharpest and contains only the modes expected[24] for the pyrochlore phase. Thus, we fitted the Raman spectrum of LZ-1450 with six Lorentzian functions for pyrochlore-related modes. In the fitting procedure we kept peak position, width and intensity of the peaks as free fitting parameters. The pyrochlore-related modes, thus obtained, are shown by filled green area, at 298 $cm^{-1}$, 315, 395 $cm^{-1}$, 503 $cm^{-1}$, 527 $cm^{-1}$ and 585 $cm^{-1}$ which matches quite well with earlier reports[9]. Figure 4 is the contour plot, demonstrating the evolution of Raman spectral profile (with $\lambda_{ex}$=488 nm) with calcination temperature, $T_c$. To show the evolution of scattering intensity of the spectral features, we plotted the intensity on a color scale. In the contour plot we find the onset of spectral line features of the pyrochlore phase at 395 $cm^{-1}$, 503 $cm^{-1}$ and 527 $cm^{-1}$ over the uniform background above 950$^o$C [white dot-dashed lines are guide to the eyes]. The appearance of these relatively sharp features along with broad mode between ~300 $cm^{-1}$−350 $cm^{-1}$ leads us to deconvolute each spectrum in Figure 3 [for LZ-850 to LZ-1100] with seven Lorentzian functions, one for the fluorite phase and six for the pyrochlore phase. In the fitting procedure the relative intensity ratios of the pyrochlore-related peaks had been kept nearly same as obtained for LZ-1450 (having only the pyrochlore phase). In Figure 3, the deconvoluted component of the fluorite phase at 350 $cm^{-1}$ is shown by the filled magenta area. We also find that the Raman spectrum of LZ-1200 is mostly dominated by the pyrochlore-related peaks. It was not possible to assign the low wavenumber Raman mode at 145 $cm^{-1}$ in LZ-750 –LZ-1000 unambiguously. Possibilities may include one of the disorder-activated Raman forbidden $F_{1U}$ modes—this is supported by the fact that the peak vanishes in case of LZ-1100 as well as for the samples calcined at higher temperatures, which reveal less-defected structures.[24]

To investigate the evolution of the relative amount of pyrochlore phase with $T_c$, we estimate the intensity ratio $R=I_{350}/I_{300}$ of all spectra. $I_{300}$ and $I_{350}$ are the intensities of the main peak of the pyrochlore and fluorite phases at 300 cm$^{-1}$ and 350 cm$^{-1}$, respectively. The value of R does not provide the exact pyrochlore to fluorite phase ratio in the system, but it yield the trend in evolution of pyrochlore fraction with $T_c$. Estimated R for LZ-750 to LZ-1450, as could be obtained from the analysis of the Raman spectra in Figure 3, is shown by open red symbols in Figure 5(a). The change in Raman shift of the most intense pyrochlore peak at 300 cm$^{-1}$, $\omega_P^{300}$, with increase in $T_c$ is shown in Figure 5(b). We find a relatively rapid increase in the peak position till LZ-1000 compared to what observed for further increase in $T_c$. Here we would like to mention that other Raman modes are either weak or strongly overlapped. The peak position obtained from the deconvolution of the spectra strongly depends on other fitting parameters. Hence, here we focus on the sharpest peak, which could be analyzed unambiguously.

Intriguing results could be obtained in Raman spectra of the LZ-1450 recorded with different excitation wavelength. Figure 6(a) compares the room temperature Raman spectra of LZ-1450 recorded with $\lambda_{ex}$=488 nm and $\lambda_{ex}$=785 nm as excitation sources. We clearly observe two peaks (marked by A and B) of very similar spectral profile at 240 cm$^{-1}$ and 360 cm$^{-1}$ in the spectrum recorded with $\lambda_{ex}$=785 nm, which were absent in the spectrum with $\lambda_{ex}$=488 nm. To find the origin of these peaks, we carried out temperature dependent Raman measurements over the temperature range between 80K and 300K [Figure 6(b)]. It is to be noted that the Raman intensity of a crystalline mode is expected to increase with a decrease in temperature[25] We clearly observe that the peaks A and B appear only upto 200K and their intensities gradually increase with temperature. Below 200K spectra carry the signatures of the pure pyrochlore phase. We find that the changes in the spectrum are reversible, i.e., both these additional peaks reappear when the temperature is brought back to 300K (topmost spectrum in Figure 6(b) shown in green color). The increase in logarithmic relative intensity of these peaks with respect to the main pyrochlore peak($I_{240}/I_{300}$ and $I_{360}/I_{300}$) with temperature are plotted in Figure 7.

**Discussion**

XRD patterns in Figure 1 and Raman spectra with $\lambda_{ex}$=488 nm in Figure 3 reveal a gradual phase transformation of disordered fluorite to pyrochlore phase with the increase in calcination temperature, $T_c$. We find that disordered fluorite phase dominates in LZ-750 (bottom panel of Figure 3). The sample LZ-1000 contains both fluorite and pyrochlore phase, whereas in the sample LZ-1450 only the pyrochlore phase can be detected by the means of XRD and Raman measurements (refer to Table 1 and Figures 2 and 3). Interestingly, with an increase in $T_c$, the main Raman mode of the pyrochlore phase at ~296 cm$^{-1}$ shifts by 4 cm$^{-1}$ (to 300 cm$^{-1}$). In the literature, this particular mode is assigned to O−Zr−O vibration in the LZ system.[8] The shift towards higher wavenumber is a clear signature of difference in oxygen environment of Zr ions with an increase in $T_c$. The possible reasons for the increase in Raman shift can be (i) a change of Zr−O bond length due to modification of atomic arrangement in the crystal structure, (ii) a change in coordination number of the cation, and/or (iii) a structural phase transition in the system. From the Rietveld analysis (refer to Table 1) we find that Zr−O bond length is slightly higher in LZ-1450 than in LZ-1000. This would have decreased the Raman shift in the system[26], in contrast to what we observe experimentally. Moreover, the x-ray data analysis shows that the sum of bond valences are nearly same in LZ-1000 and LZ-1450. Thus, we believe that the rearrangement in anionic sublattice in this system during structural phase transition from the disordered fluorite to pyrochlore phase results in the shift in pyrochlore-related peak, as observed in Fig. 5(b). Above conjecture of the deviation of perfect oxygen environment in atomic arrangement of the pyrochlore phase of $La_2Zr_2O_7$, is further supported when we study the Raman spectra of LZ-1450 recorded with $\lambda_{ex}$=785 nm. It was interesting to find additional Raman spectral features A and B in this sample for $\lambda_{ex}$=785 nm, which are clearly absent in the spectrum recorded with $\lambda_{ex}$=488 nm (refer to Figure 6(a)).

In a dielectric medium the amplitude of electromagnetic radiation decays exponentially along its path. The penetration depth (δ) of a particular wavelength in a medium is the measure over which the amplitude of radiation decays 1/e times of its original value and is proportional to the wavelength *(λ)* of incident radiation by $\delta \equiv \frac{\lambda}{Im(n(\lambda))}$. Im *n(λ)* is the complex part of the refractive index of the material[27]. Thus, the depth of penetration for 785 nm laser radiation is expected to be 1.6 times larger than that of 488 nm, if we assume *n(λ)* does not vary appreciably for these two wavelengths. We also know that x-ray penetrates deep inside the bulk and is sensitive to the long range ordering in the crystal structure. From above mentioned considerations it is reasonable to believe that the additional peaks in the spectra of LZ-1450 with $\lambda_{ex}$=785 nm carries the information of local crystal structure in the bulk, which were absent either in long range ordering of atoms or near the surface region. In the literature the difference in penetration depth of electromagnetic radiation has been exploited to study the structural phase transition in a compound[28,29], For example, UV and visible Raman measurements could reveal the difference in structural phase in the near surface and bulk region of $TiO_2$.[28]

Next, we try to study the origin of these additional Raman peaks. We find both these spectral profiles are identical in nature, i.e., they have similar spectral line profile. These peaks are shifted by 60 cm$^{-1}$ on the both sides of the intense pyrochlore peak at 300 cm$^{-1}$. Moreover, the temperature variations of both these spectral features are very similar, as shown in Figure 6(b) and 7. As mentioned earlier, for a fluorite structure we expect only one Raman mode. Thus, the appearance of two identical peaks rules out the possibility of the existence of local fluorite related phase in the crystal structure of LZ-1450. Assuming the origin of these peaks to be local disorders in the crystal structure, we fitted the variation of logarithmic relative intensity of these peaks with temperature (in Kelvin) using the equation, $I=I_0e^{-E_a/KT}$, where, $I_0$ is a constant and $E_a$ is related to the activation energy of the defect states related to these peaks A and B. We find the values of $E_a^A$ =0.05±0.01eV and $E_a^B$ =0.06±0.01 eV. These values are very close to the activation energy for thermal annealing of the disordered states in zirconium pyrochlores.[27] The absence of these peaks in the near surface region, furthermore, suggests that they are associated with the oxygen related disordered states in the pyrochlore phase.

Most of the experiments, e.g., XRD measurements, Raman spectroscopy or thermal calorimetry, which have been extensively used to study to the phase transition in lanthanum zirconate, look into bulk behaviour under different experimental conditions. If oxygen be one of the factors determining the order-disorder phase transition in a system, high temperature calorimetric measurements in the air is, indeed, not expected to follow the experimental results obtained at ambient condition, for example, by Raman or XRD measurements. The tests carried out via differential thermal calorimetry (DSC) focus on a dynamic situation where role played by the local defects are rarely understood. Moreover, DSC is not the most reliable technique for observing second order phase transitions, i.e., when no sharp change in enthalpy is observed. As the crystal structure of fluorite and pyrochlore phase are closely related, these oxygen related disordered states may play crucial role in the phase transition of the system. The oxygen-related defect states are also expected to play a significant role for the related functional properties of the system and, therefore, will be investigated in future.

**Conclusion**

$La_2Zr_2O_7$ powders synthesised by co-precipitation was investigated in detail for the fluorite-pyrochlore phase transition behaviour below 1500 °C. Our Rietveld and Raman measurements address the long standing debate on phase transition in this system. X-ray diffraction data analysis reveals that there is a disordered fluorite to pyrochlore transition in $La_2Zr_2O_7$ between 1000 °C and 1450°C. However, we claim that the dynamics of the phase transition is different in near-surface and bulk. We report the existence of local oxygen-related disordered states in the bulk of pyrochlore LZ system, which are absent in its near surface regions. It was found that the activation energy of these states is very close to the thermal energy at room temperature. Below 200K these states disappear and a pure pyrochlrore LZ sysem is obtained. We believe that these states may play a crucial role in determining the parameters for fluorite to pyrochlore phase transition in $La_2Zr_2O_7$.

**Acknowledgements**

AR thanks BRNS, India for financial assistance

***Corresponding author**: fax: +91-612-2277383.
E-mail address: anirban.chowdhury@gmail.com.

**Author Contributions**

The manuscript was written through contributions of all authors. All authors have given approval to the final version of the manuscript. All authors contributed equally.

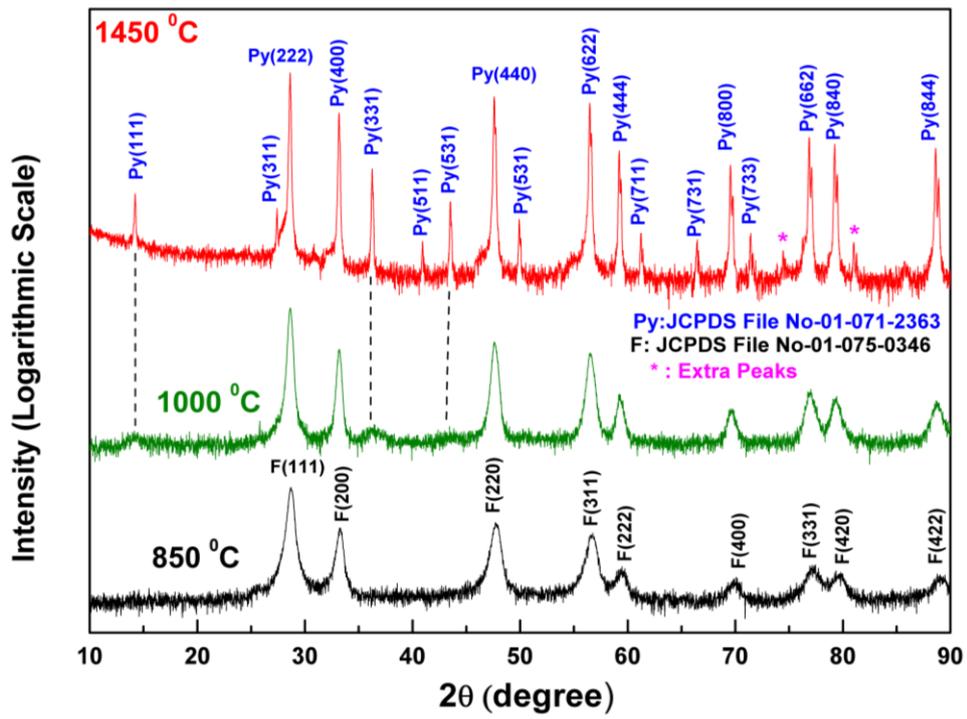

**Figure 1**

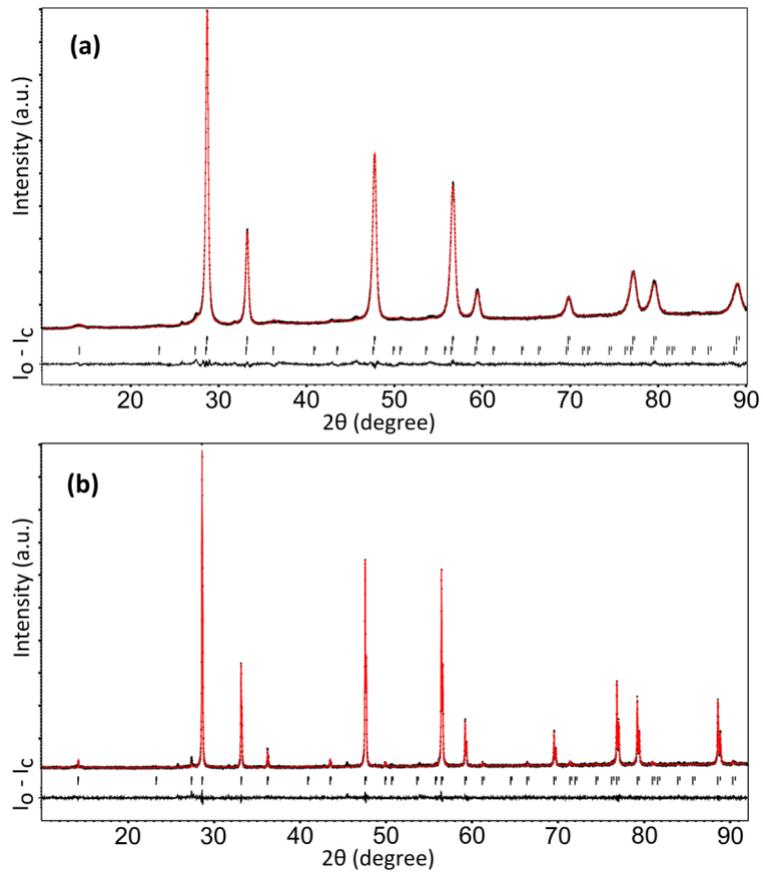

**Figure 2**

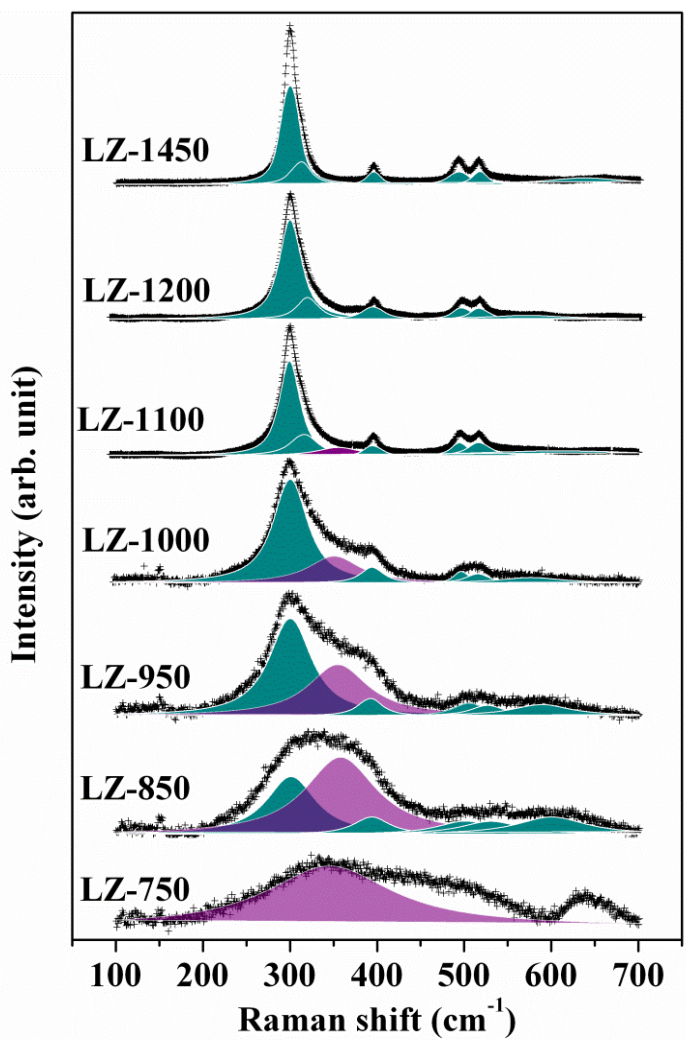

**Figure 3**

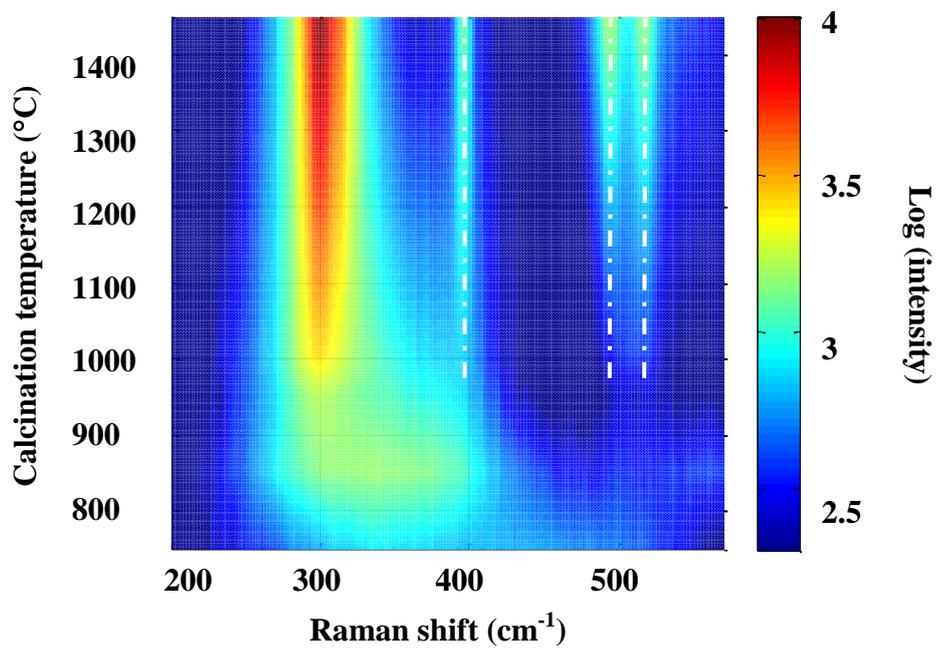

**Figure 4**

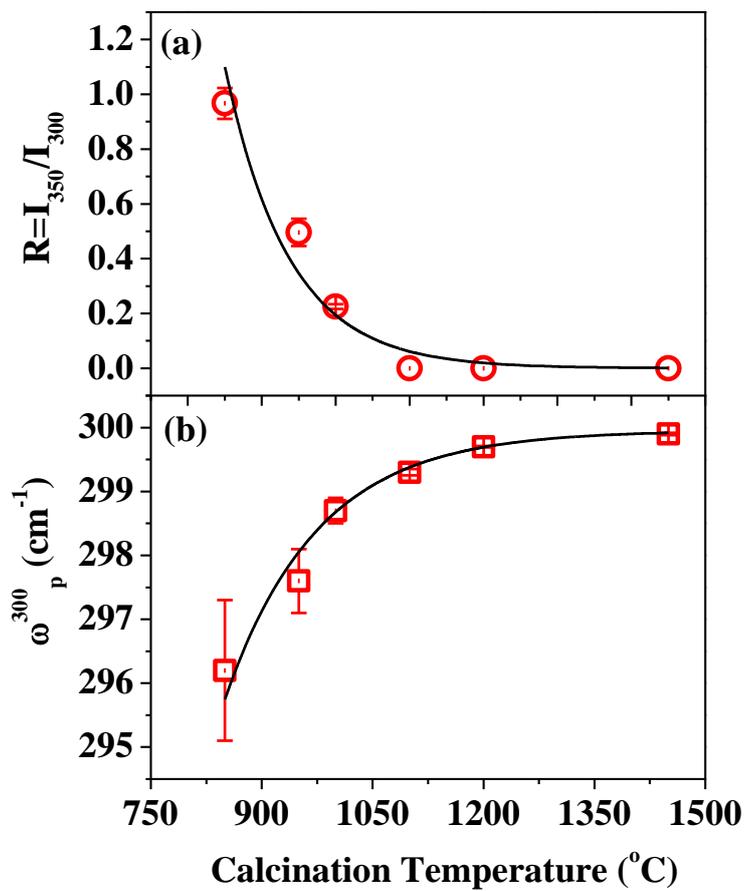

**Figure 5**

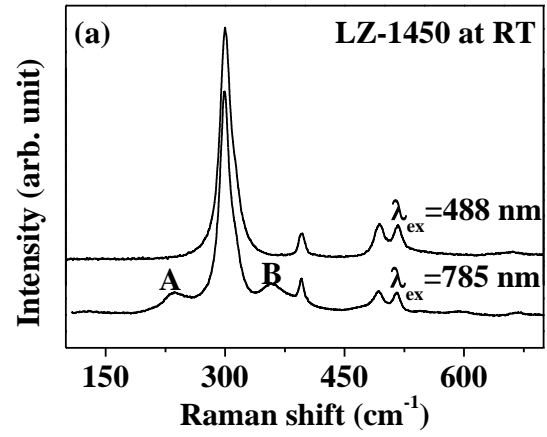

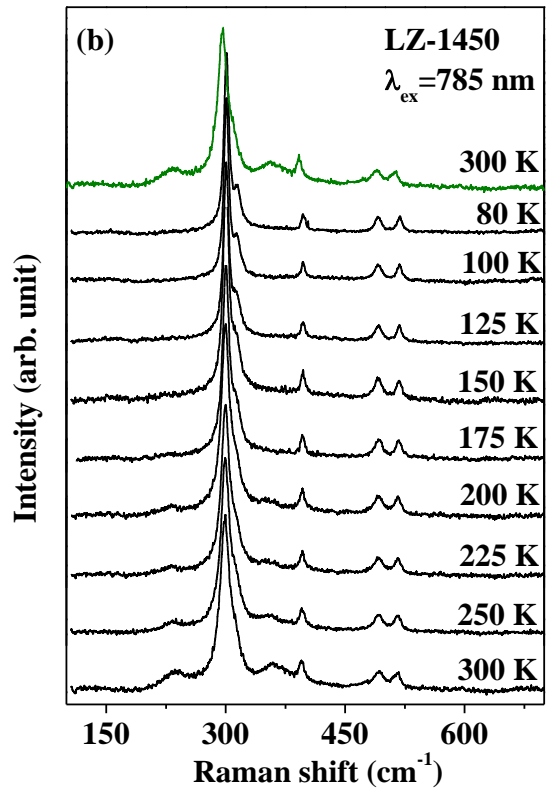

**Figure 6**

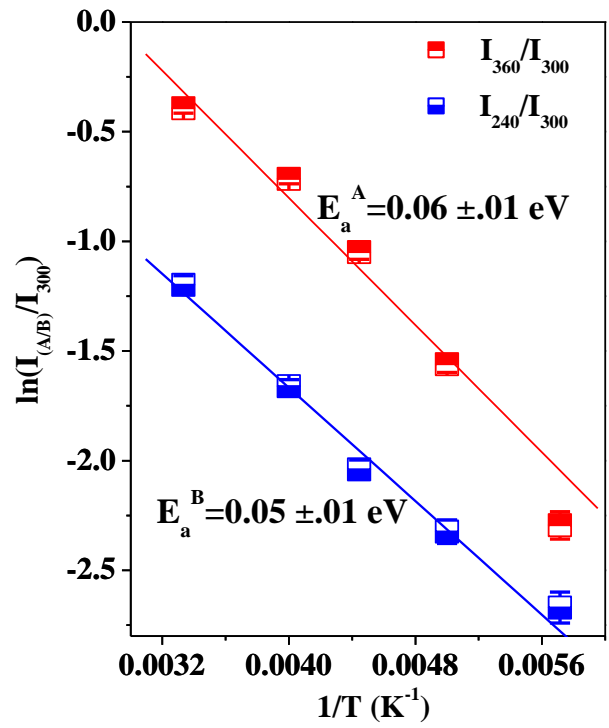

**Figure 7**

**List of Figures**

Figure 1: XRD patterns for LZ samples calcined at different temperatures; the dotted lines denote the peaks exclusively occurring for a pyrochlore phase.

Figure 2: Results of Rietveld refinement of the samples LZ1000 and LZ-1450. The observed (·) and calculated (−) powder XRD patterns have been plotted, with the difference curve in the bottom of each figure.

Figure 3: Room temperature Raman spectra of samples calcined at different temperatures (750, 950, 1000, 1100, 1200 and 1450°C). The solid lines are the net fitted spectrum to the data points shown in + symbols. Pyrochlore modes are shown with filled green and fluorite with filled magenta color.

Figure 4: Contour plot of the $T_c$ dependence of the Raman spectral profile over the above spectral between 200 and 600 cm$^{-1}$. The dashed lines mark the signature of appearance sharper spectral features above 950°C.

Figure 5: Evolution of (a) Raman intensity $R=I_{300}/I_{350}$ and (b) Raman shift of the pyrochlore mode, $\omega_P^{300}$, as a function of $T_c$ with $\lambda_{ex}$=488 nm

Figure 6: (a). Room temperature Raman spectra of LZ-1450 recorded with $\lambda_{ex}$=488 nm and $\lambda_{ex}$=785 nm. A and B mark additional Raman peaks observed with $\lambda_{ex}$=785 nm. (b) Temperature dependent Raman measurements on LZ-1450 with $\lambda_{ex}$=785 nm.

Figure 7: ln($I/I_{300}$) vs T plot for the peak A (blue) and B (red).

**Table I. Refined unit cell parameter and bond lengths in LZ-1000 and LZ-1450**

| Phase | Sample | LZ-1000 | LZ-1450 |
|---|---|---|---|
| Fluorite $La_{0.5}Zr_{0.5}O_{1.75}$ | Content [wt%] | 46(5) | 0 |
| | a [Å] | 5.385(2) | - |
| | La/Zr – O [Å] | 2.3316(10) | - |
| | La – Zr [Å] | 3.807(2) | - |
| Pyrochlore $La_2Zr_2O_7$ | Content [wt%] | 54(5) | 100 |
| | a [Å] | 10.799(6) | 10.798(3) |
| | La – O1 [Å] | 2.6411(17) | 2.63(3) |
| | La – O2 [Å] | 2.3381(14) | 2.3379(9) |
| | Zr – O1 [Å] | 2.0999(14) | 2.105(18) |
| | La – Zr [Å] | 3.818(3) | 3.8178(13) |
| Profile parameters | wRp [%] | 3.36 | 9.70 |
| | GOF | 1.58 | 1.04 |